\documentstyle[psfig,epsfig]{mn}
\pssilent 

\def\Mpc{\ifmmode {\, h^{-1} \, {\rm Mpc}}
\else {$h^{-1}\,$ Mpc}\fi}

\def\s8{{\sigma_8}}

\def\ltsima{$\; \buildrel < \over \sim \;$}
\def\simlt{\lower.5ex\hbox{\ltsima}} 
\def\gtsima{$\; \buildrel > \over \sim \;$} 
\def\simgt{\lower.5ex\hbox{\gtsima}} 
\def\bfw{{\bf w}}

\def\omegab{{\Omega_{\rm b}}}

\def\omegabh2{{\omegab h^2}}

\begin{document}

\title[Joint estimation with 'hyper-parameters']
{Bayesian `Hyper-Parameters' Approach 
to Joint Estimation:
The Hubble Constant from CMB Measurements} 
\author[O.~Lahav et al.]
{O. Lahav$^{1,2}$, S.L. Bridle$^{3}$, 
M.P. Hobson $^{3}$, A.N. Lasenby$^{3}$, \and 
\& L. Sodr\'e Jr.$^{4}$ \\
$1$ Institute of Astronomy, Madingley Road, Cambridge CB3 0HA; email: lahav@ast.cam.ac.uk \\ 
$2$ Racah Institute of Physics, The Hebrew University, Jerusalem 91904, Israel \\
$3$ Astrophysics Group, Cavendish Laboratory, Madingley Road, Cambridge CB3 0HE\\
$4$ Departamento de Astronomia, Instituto Astronomico e Geofisico 
da USP, Av Miguel Stefano 4200, 04301-904 S\~ao Paulo, Brazil \\
}

\maketitle

\begin{abstract}
Recently  several studies have jointly analysed data from
different  cosmological probes with the motivation of estimating 
cosmological parameters.
Here we generalise this procedure to take into account the relative
weights of various probes. This is done by including in the joint 
$\chi^2$ function a set of 
`Hyper-Parameters', which are dealt with using  Bayesian considerations.
The resulting algorithm 
(in the case of uniform priors on the log of the Hyper-Parameters) 
is very simple: instead of minimising 
$\sum \chi_j^2$ 
(where $\chi_j^2$ is per data set $j$) 
we propose to minimise  $\sum N_j \ln (\chi_j^2)$ 
(where $N_j$ is the number of data points per data set $j$). 
We illustrate the method by  estimating  the Hubble constant $H_{0}$ from 
different sets of recent CMB experiments (including Saskatoon, Python V, 
MSAM1, TOCO and Boomerang).
\end{abstract}

\section{Introduction}
\label{intro}

Several groups  (e.g. Eisenstein et al. 1999, Gawiser \& Silk 1998, 
Bridle et al. 1999, Bahcall et al. 1999, 
Bond \& Jaffe 1998, Lineweaver 1998) 
have recently estimated cosmological parameters by joint analysis 
of data sets (e.g. CMB, SNe Ia, redshift surveys, cluster abundance
and peculiar velocities).

A complication that arises in combining data sets 
is that there is freedom
in assigning  the relative weights of different measurements.
Some approaches to this problem 
have been suggested in the astronomical literature
(e.g. Godwin \& Lynden-Bell 1987; Press 1996).  
Here we propose a Bayesian approach utilizing `Hyper Parameters'
(hereafter HPs).

Assume that we have 2 independent data sets, $D_{A}$ and $D_{B}$
(with $N_{A}$ and $N_{B}$ data points respectively) 
and that we wish to determine a vector of free parameters ${\bfw}$
(such as the density parameter $\Omega_{\rm{m}}$, the Hubble constant $H_0$ etc.).
This is commonly done by minimising  
\begin{equation}
\chi^2_{\rm joint} = \chi^2_A \; + \; \chi^2_B\; , 
\label{chi2_simple}
\end{equation}
(or maximizing the sum of log  
likelihood functions).
Such procedures assume that the quoted observational random errors 
can be trusted, and that the two (or more) $\chi^2$s  
have equal weights.  
However, when combining `apples and oranges' one may wish to allow freedom in 
the relative weights. 
One possible approach is to generalise Eq. \ref{chi2_simple} to be 
\begin{equation}
\chi^2_{\rm joint} = \alpha  \chi^2_A \; + 
                \beta \; \chi^2_B \; , 
\label{chi2_hp}
\end{equation}
where $\alpha$ and $\beta$ are
`Lagrange multipliers', or `Hyper-Parameters',
which are   to be evaluated 
in a Bayesian way.
There are a number of ways to interpret the meaning of the HPs.
A simple example of the HPs is the case
that 
\begin{equation}
\chi^2_{A} = 
\sum {\frac {1} {\sigma_{i}^{2}}}  [x_{{\rm obs},i} - x_{{\rm pred},i}(\bfw)]^2\;,
\label{chi2_example}
\end{equation}
where the sum is over $N_{A}$ 
measurements and corresponding predictions and errors
$\sigma_{i}$. Hence by multiplying $\chi^2$ by $\alpha$
each  error 
effectively 
becomes ${\alpha}^{-1/2} \sigma_{i}$.
But even if the measurement errors are accurate, 
the HPs are useful in assessing the relative weight of different experiments.
It is not uncommon that astronomers 
discard measurements (i.e. by assigning $\alpha=0$)
in an ad-hoc way. The procedure we propose gives an objective diagnostic 
as to which measurements are problematic 
and deserve further understanding of systematic or random 
errors.

We show below that 
if the prior probabilities 
for $\ln (\alpha) $ and $\ln (\beta) $ are uniform then 
one should consider the quantity 
\begin{equation}
-2 \; \ln P(\bfw| D_{A}, D_{B}) =   N_{A} \ln 
(\chi_{A}^{2})   + N_{B} \ln 
(\chi_{B}^{2}) 
\label{chi2_new}
\end{equation} 
instead of Eq. \ref{chi2_simple}.
It is as easy to calculate this statistic as the standard $\chi^2$.
The effective HPs can then be identified as  
$\alpha_{\rm eff} = \frac { N_{A}} {\chi_{A}^{2}}$ 
and
$\beta_{\rm eff} = \frac { N_{B}} {\chi_{B}^{2}}$,
where $\chi^2_A$ 
and $\chi^2_B$ are computed at the values of the parameters 
${\bf w}$ that 
minimise eq. \ref{chi2_new}.
The derivation and interpretation of these results are given 
in Section 2. In Section 3 we apply the method to a set of CMB 
experiments and estimate the best fit Hubble constant.
Extensions of the methods are discussed in Section 4. 

\section{`Hyper-parameters'}

How do we eliminate the unknown HPs  $\alpha$ and $\beta$ ?
We follow here the Bayesian formalism given in Gull (1989), MacKay (1994) 
and Bishop (1995). The formalism in these references was given in the 
context of Maximum Entropy and Artificial Neural Networks.

By marginalisation over $\alpha$ and $\beta$ we can write
the probability for the parameters $\bfw$ given the data:
\begin{equation}
P(\bfw| D_{A}, D_{B}) = \int \int P(\bfw, \alpha, \beta | D_{A}, 
D_{B})  \; d \alpha \;d \beta \;.
\end{equation}
Using Bayes' theorem we can write the following relations:
\begin{equation}
P(\bfw, \alpha, \beta | D_{A}, D_{B}) = 
\frac {P(D_{A}, D_{B} | \bfw, \alpha, \beta) \;P(\bfw, \alpha, \beta) } 
{P(D_{A}, D_{B})} \;,
\end{equation} 
and 
\begin{equation}
P(\bfw, \alpha, \beta) = P(\bfw | \alpha, \beta) \; P(\alpha, \beta) \;. 
\end{equation}

We now make the following assumptions:
\begin{equation}
P( D_{A},D_{B}|\bfw, \alpha, \beta) = 
P(D_{A}| \bfw,\alpha) \; P(D_{B}|\bfw,\beta) \;,
\end{equation} 

\begin{equation}
P(\bfw |\alpha, \beta) = const. \;,
\end{equation}

\begin{equation}
P(\alpha, \beta) = P(\alpha) \; P(\beta) \;.
\end{equation} 
With the choice of `non-informative' uniform priors in the log, 
$P(\ln \alpha) = P(\ln \beta) =1$ (Jeffreys 1939) we get
$P(\alpha) = 1/\alpha$  and $P(\beta) = 1/\beta$.
Note that the integral over priors of this kind
diverges
(such a prior is called `improper', see Bishop 1995).
These are very conservative prior, essentially 
stating that we are ignorant about the 
scale of measurements and  errors.
The other extreme is obviously $P(\alpha) = \delta(\alpha-1)$, 
i.e. when the measurements and  errors are taken faithfully. 
One can try other forms (see below), 
but it is likely that these 2 extreme forms 
reasonably bracket the probability space.  
Hence:
\begin{equation}
P(\bfw| D_{A}, D_{B}) =  { 1 \over P(D_{A}, D_{B}) } \; P(D_{A} | \bfw) 
\;P(D_{B} | \bfw), 
\label{PwDADB}
\end{equation} 
where 
\begin{equation}
P(D_{A} | \bfw) \equiv \int P(D_{A} | \bfw, \alpha) \alpha^{-1} d \alpha 
\;,
\label{PDA}
\end{equation} 
and
\begin{equation}
P(D_{B} | \bfw) \equiv \int P(D_{B} | \bfw, \beta)  \beta^{-1} d \beta \;.
\label{PDB}
\end{equation}

 It is common to have a  
 likelihood function of the form of a Gaussian in $N_{A}$ 
 dimensions:
 \begin{equation}
 P_{G}(D_{A}|\bfw) \; \propto \;\exp[ -\chi_{A}^2 /2]\;,  
\label{mvGauss}
\end{equation}
where we assume  
for simplicity that the normalization 
constant is independent  of the parameters
$\bfw$ (this is indeed the case in our application for the CMB 
measurements in the next Section).


We generalise this form to incorporate $\alpha$ as follows:
\begin{equation}
P(D_{A} | \bfw, \alpha) \; \propto \; \alpha^{N_{A}/2}\; 
\exp (-{\alpha \over 2} \chi_{A}^{2} ) \; .
\label{Gauss_alpha}
\end{equation} 
The integral of Eq. \ref{PDA} then gives 
\begin{equation}
P(D_{A} | \bfw) \propto  
{(\chi_{A})}^{-N_{A}}\;,
\label{power-law}
\end{equation}
and similarly for Eq. \ref{PDB}.
We note that it is the specific choice of prior for $P(\alpha)=1/\alpha$
that has led to a change from a  Gaussian distribution (Eq. \ref{mvGauss} )
to a power-law (Eq. \ref{power-law}).

Eq. \ref {PwDADB} can then be written (ignoring constants) as
\begin{equation}
-2 \; \ln P(\bfw| D_{A}, D_{B})\; = 
N_{A} \ln (\chi_{A}^{2})   
\; + \;   N_{B} \ln (\chi_{B}^{2})\;. 
\label{lnPwDADB}
\end{equation} 
To find the  best fit parameters $\bfw$ requires us to minimise
the above probability in the $\bfw$ space.
Eq. \ref{lnPwDADB} generalises a similar
equation derived by Cash (1979).
Cash used a 
very different  set of arguments based on 
maximum likelihood 
and he assumed that the error per group of data is the same
(in this special case the  
original quoted errors drop out in the minimisation).
We emphasize that
our Bayesian framework is more general and `principled', 
and therefore we can derive alternative
equations by assuming different priors.

Since $\alpha$ and $\beta$
have been eliminated from the analysis by
marginalisation they do not have particular values that can be quoted.
Rather, each value of $\alpha$ and $\beta$ has been considered, and weighted
according to the probability of the data given the model.
However, it may be useful to know which
values of $\alpha$ and $\beta$
were given the most weight. This can be
estimated by finding the values of 
$\alpha$ and $\beta$ at which 
eq \ref{Gauss_alpha} peaks:
\begin{equation}
\alpha_{\rm {eff}} = \frac { N_{A}} {\chi_{A}^{2}}\;,  
\label{alpha1}
\end{equation} 
and similarly
\begin{equation}
\beta_{\rm {eff}} = \frac { N_{B}} {\chi_{B}^{2}} \;, 
\label{beta1}
\end{equation} 
both evaluated at the joint peak.

We note that if we substitute these effective $\alpha$ and $\beta$
in Eq. \ref{chi2_hp} we obtain $\chi^{2}_{joint} = N_{A} + N_{B}$, 
i.e. a reduced $\chi^2$ of unity 
(for the  case when the
number of degrees of freedom is dominated by the number of data points).

There is of course freedom in choosing the prior. 
For example, if we take $P(\alpha) =1 $ (instead of Jeffreys' prior
$P(\alpha) = 1/\alpha$) we find that the function to be minimised is
$(N_A + 2) \ln (\chi^2_{A})$,
instead of  
$N_A \ln (\chi^2_{A})$.
Thus these two priors give very similar results for large $N_A$.
Numerous other priors are possible (e.g. a top-hat centred on 
a plausible value), but at the expense of more free HPs
(e.g. the width of the top-hat).
 
\section {Application to CMB data}

We illustrate the  effect of using HPs by application to
measurements of the angular power spectrum of the cosmic microwave
background (CMB). Numerous groups have now used CMB data to
estimate cosmological parameters (see Rocha 1999 for a review). 
The most common method is the flat bandpower method (Bond
1995a; Bond 1995b) in which the difference between observed and
predicted flat bandpowers are compared using the
$\chi^2$ statistic (Eq. \ref{chi2_example}). 
There is a question as to whether one should use all measurements from
all groups of observers, independent of whether a given data set is
consistent with the other data. 
Dodelson and Knox (1999) do address
this issue by assigning a calibration coefficient to each data
point, the values of which are optimised for each cosmological model
investigated. The HPs method offers a Bayesian alternative
to ad-hoc selection of data sets or the problems associated with using
incompatible data sets and a conventional approach.

\begin{figure}
\centerline{\vbox{
\epsfig{file=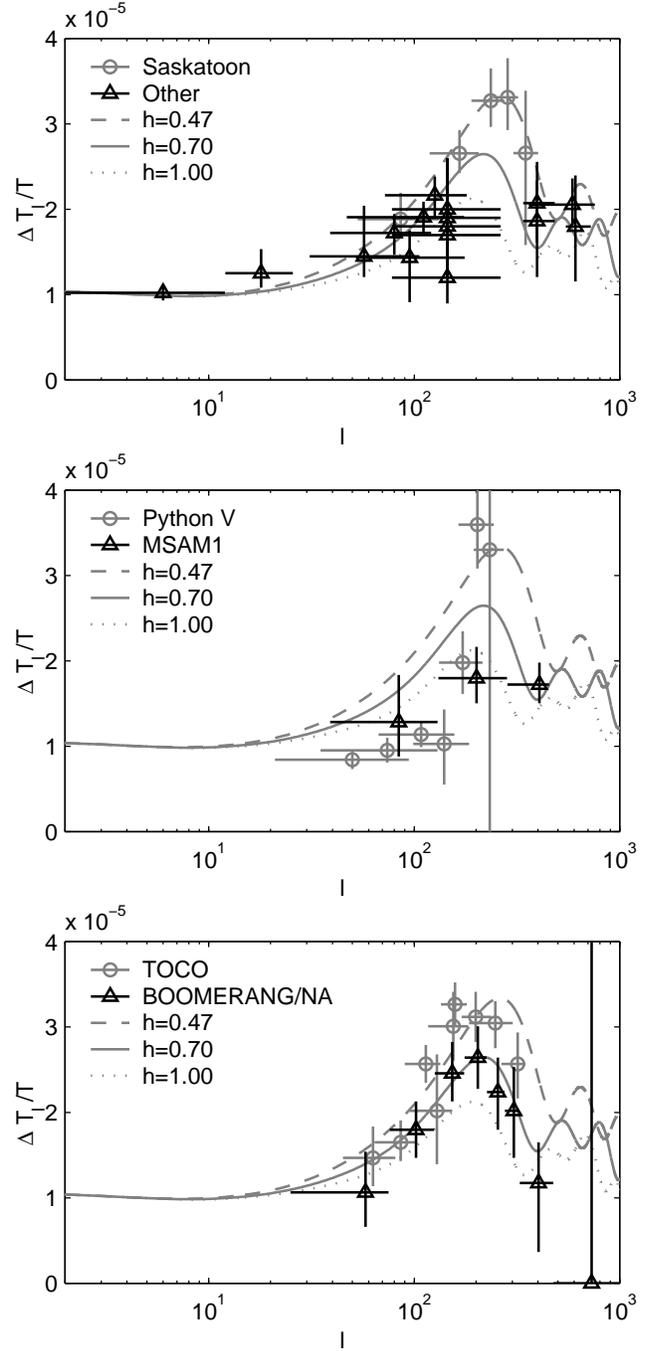,width=8.5cm}
}}
\caption{The CMB data used, shown in three panels for clarity.
Theoretical CMB power spectra are shown for $h=0.47$, $h=0.70$ and
$h=1.00$ (other parameters are fixed at the values given in the text).
\label{cmbdata}}
\end{figure}

There are clearly a large number of possible combinations of CMB data
sets that could be investigated. For the purpose of illustration we
divide a selection of the current CMB power spectrum estimates into
six subsets.
The subsets are (i) Saskatoon (Netterfield et al. 1997,
including the five per cent calibration error; Leitch, 
private communication), (ii) Python V (Coble
et al. 1999), (iii) MSAM1 (Wilson et al 1999), (iv) TOCO (Torbet et al. 1999,
Miller et al. 1999), (v) BOOMERANG/NA (Mauskopf et al. 1999, assuming
Gaussian window functions which fall by a factor of $1/e$ at
$\ell_{\rm{min}}$ and $\ell_{\rm{max}}$ as specified in the paper)
and (vi) we group all of the remaining
data into the fifth subset and refer to it as `Other'. This subset
contains COBE, Tenerife, South Pole, ARGO, MAX, QMAP, OVRO and CAT (see
Hancock et al. 1998, Webster et al. 1998 and Efstathiou et al. 1999
for more details).
These data are plotted in Fig.~\ref{cmbdata}.
\begin{figure}
\centerline{\vbox{
\epsfig{file=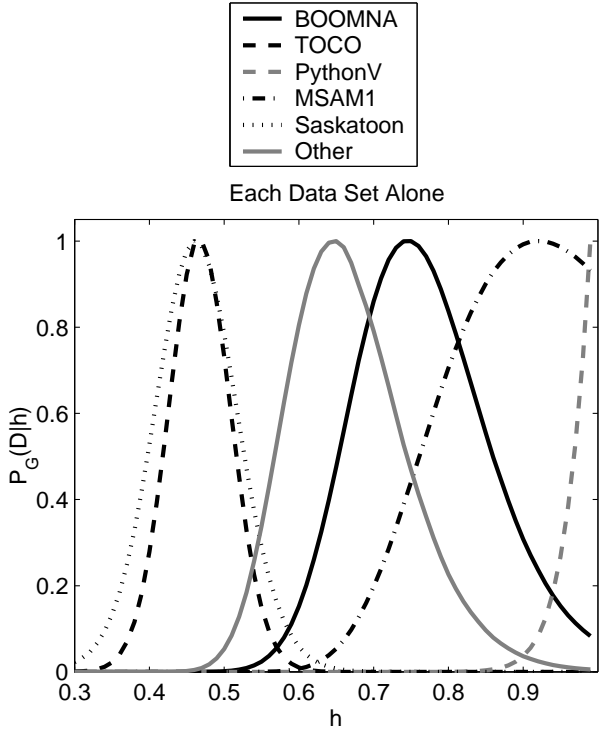,width=8cm}
}}
\label{eachalone}
\caption{ The probability of the Hubble constant $h$ as a function of
$h$ from different subsets of CMB data (as indicated in the legend)
resulting from a conventional $\chi^2$ analysis.
}
\end{figure}
\begin{table}
\center{\vbox{
\begin{tabular}{@{}lrrr}
\hline
Data         & $N_j$ & Best $h$ &   $\chi^2_j$ \\
\hline
BOOMERANG/NA & 7     & $0.75$ & 1.5 \\
\hline
TOCO         & 9     & $0.47$ & 10.5 \\
\hline
MSAM1        & 3     & $0.92$ & 2.1 \\
\hline
PythonV      & 7     & $1.00$ & 48.3 \\
\hline
Saskatoon    & 5     & $0.46$ & 0.5 \\
\hline
Other        & 16    & $0.65$ & 16.3 \\
\hline
\end{tabular}
}}
\label{tableeachalone}
\caption{
Conventional $\chi^2$ analysis using each data subset alone. 
For each data subset the number of data points, $N_j$, the best fit
value of $h$ and the $\chi^2$ value at this point is shown. The full
likelihood distributions in $h$ are shown in Figure \ref{eachalone}.
}
\end{table}

In addition, for simplicity we restrict ourselves to a very limited
set of cosmological models. We assume CMB fluctuations arise from
adiabatic initial conditions with cold dark matter and negligible
tensor component, and that $\Omega_{\rm{m}}=0.3$,
$\Omega_{\Lambda}=1-\Omega_{\rm{m}}=0.7$, $n=1$,
$Q_{\rm{rms-ps}}=18\mu$K and $\Omega_b h^2 = 0.019$. We then
investigate the constraints on the remaining parameter, the
dimensionless Hubble constant, $h=H_0/(100$~\mbox{$\rm{km} \rm{s}^{-1}
\rm{Mpc}^{-1}$}$)$. Theoretical power spectra for three
different values of $h$ are shown in Fig.~\ref{cmbdata}: increasing
$h$ decreases the height of the first acoustic peak, and makes few
other significant changes for the purpose of our analysis. 
The range
in $h$ investigated here ($0.3<h<1.0$) takes the peak height from
above the Saskatoon upper error bars down to the MSAM1 points.

To aid qualitative understanding of the analysis that follows, it is
helpful to first calculate the value of $h$ preferred by each data
subset. The results are plotted in Fig. 2 
and shown in Table 1.
As expected from the range of first acoustic peak heights
preferred by the data, the $h$ values also vary considerably.
The BOOMERANG/NA data alone prefers an intermediate value of $h$, as
does the `Other' data. TOCO and Saskatoon both agree on a relatively
high first acoustic peak and so on a low $h$. The MSAM1 points are
quite low and thus fit a high value of $h$, and the PythonV points
also prefer a high value of $h$, although the $\chi^2$ value is very
high, indicating a bad fit to the model.

\begin{figure}
\centerline{\vbox{
\epsfig{file=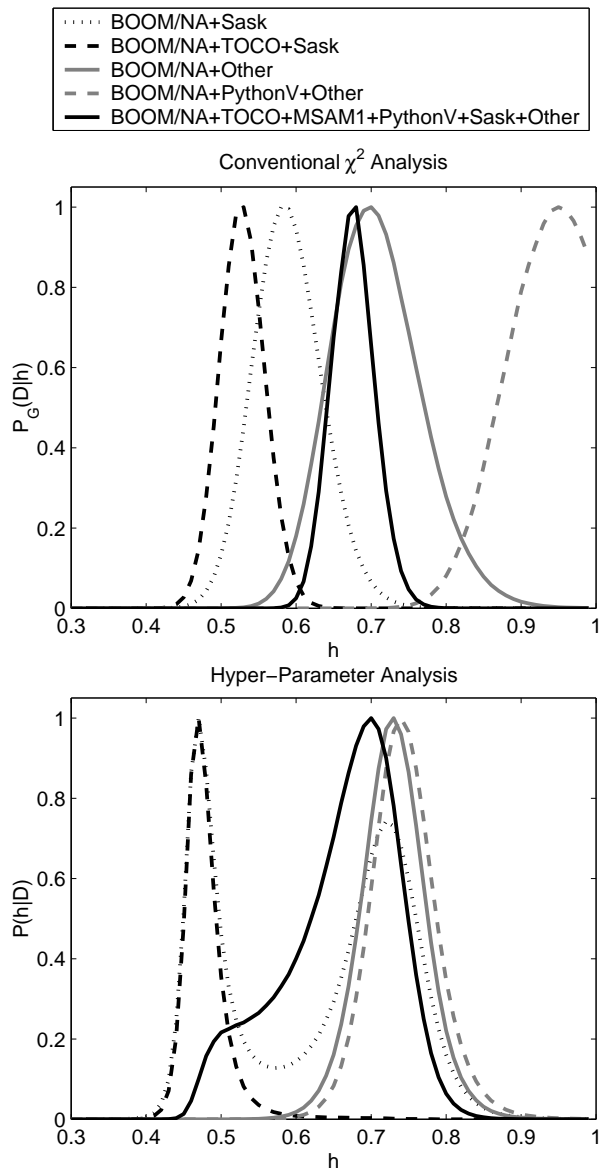,width=8cm}
}}
\caption{ The probability of the Hubble constant $h$ as a function of
$h$ from different subsets of CMB data (as indicated in the legend)
resulting (Top figure) 
from a conventional $\chi^2$ analysis and (Bottom figure) the
Hyper-Parameter analysis, in which
$P(h|D_A,D_B,\ldots)=\exp[-{\frac {1} {2}} \sum N_j \ln
(\chi^2_{j})]$.  
\label{constraints}}
\end{figure}

Clearly there are a large number of possible groupings of the data
subsets. We show here the results from just five groupings, which are
a fair sample and also highlight some of the properties of
HPs. Firstly we consider the case of two relatively discrepant data
sets, Saskatoon and BOOMERANG/NA; the $h$ values that they prefer do
not overlap significantly. Combining their $\chi^2$ values for each
$h$ in the conventional manner (Eq. \ref{chi2_simple}) yields the
likelihood function plotted with the dotted line in
Fig. \ref{constraints} (Top). An intermediate value of $h$ is preferred,
and in fact the best fitting $h$ values for each data set alone are
essentially ruled out. In contrast, when HPs are used, i.e. the
$\chi^2$ values are combined using Eq. \ref{lnPwDADB}, the dotted line
in Fig. \ref{constraints} (Bottom) is obtained. There are two peaks in the
probability distribution corresponding to the two different values of
$h$ preferred by each data set alone. This is perhaps closer to what
we would actually believe given just these two data subsets.

Next we consider the effect of adding in a data subset that agrees
strongly with one of the above two data subsets. That is, we consider
TOCO with Saskatoon and BOOMERANG/NA. The probability distribution
calculated using HPs now loses its second peak, retaining the one that
agrees with TOCO and Saskatoon. The theoretical CMB power spectrum for
the preferred value of $h=0.47$ is shown in Fig. \ref{cmbdata}.

On combining two data sets that do agree well, BOOMERANG/NA and
`Other', there is little difference between the conventional and HP
analyses, although the error bar on $h$ is slightly decreased when
using HPs.

Adding in a data subset that has a poor $\chi^2$ (given the range of
models considered), PythonV, makes a large difference to the
conventional analysis but only a very small difference to the HP
analysis. This can also be seen from the effective HP value at the
best fit $h$, calculated from Eq. \ref{alpha1}, which is much less
than unity for the PythonV data, indicating that it has been down
weighted. 

Finally we use all of the data subsets, obtaining the solid lines in
Fig. \ref{constraints}. It turns out that the best fitting value of
$h$ is similar in both the conventional and HP analyses, but the error
bars are significantly wider in the HP analysis, which corresponds
better to what we would naturally believe. 

\begin{table}
\center{\vbox{
\begin{tabular}{@{}lrrr}
\hline
Data & $N_j$ & Best $h$ &   $\chi^2_j$ \\
\hline
BOOM/NA+Sask & 12 & $0.58$ & 11.3 \\
\hline
BOOM/NA+TOCO+Sask & 21 & $0.53$ & 25.6 \\
\hline
BOOM/NA+Other & 23 & $0.70$ & 18.6 \\
\hline
BOOM/NA+PythonV+Other & 30 & $0.95$ & 81.2 \\
\hline
All data & 47 & $0.68$ & 152.6 \\
\hline
\end{tabular}
}}
\label{table}
\caption{
Conventional $\chi^2$ analysis. Best fit values of $h$ and $\chi^2$
values at this best fit point, which can be compared to the total
number of data points, $N_j$.
}
\end{table}

\begin{table}
\center{\vbox{
\begin{tabular}{@{}lrrr@{}l@{}lr@{}l@{}l}
\hline
 Data & $N_j$ & Best $h$ & $\chi^2_j$&& & Effective &&HP \\
\hline
BOOMERANG/NA    & 7  & 0.47    &  18&.&5  & 0&.&4 \\
Saskatoon       & 5  &         &  0&.&5   & 10&.&2 \\
\hline
BOOMERANG/NA    & 7  & 0.47    & 18&.&5  & 0&.&4   \\
TOCO            & 9  &         & 10&.&5  & 0&.&9 \\
Saskatoon       & 5  &         & 0&.&5   & 10&.&2  \\
\hline
BOOMERANG/NA    & 7   & 0.73   &  1&.&5  & 4&.&5  \\
Other           & 16  &        &  17&.&4 & 0&.&9\\
\hline
BOOMERANG/NA    & 7   & 0.74   &  1&.&5  & 4&.&5  \\
PythonV         & 7   &        &  73&.&1 & 0&.&1\\
Other           & 16  &        &  17&.&6 & 0&.&9\\
\hline
BOOMERANG/NA    & 7  & 0.70   &  1&.&8   &  3&.&9  \\
TOCO            & 9  &        &  34&.&9  & 0&.&3  \\
MSAM1           & 3  &        &  5&.&4   & 0&.&6\\
Python V        & 7  &        &  79&.&6  & 0&.&1\\
Saskatoon       & 5  &        &  14&.&9  & 0&.&3  \\
Other           & 16 &        &  16&.&8  & 1&.&0 \\
\hline
\end{tabular}
}}
\label{table}
\caption{
The five different combinations used in the Hyper-Parameter
analysis. In each case a separate Hyper-Parameter was given to each
data subset, the number of data points in each data subset, $N_j$, is
shown. The best fitting value of $h$, the $\chi^2$ value for each data
subset for this best fitting $h$ and the effective HP ($N_j/\chi^2_j$)
at this $h$ is calculated. 
}
\end{table}

\section {Discussion} 

We have presented a formalism for analyzing a set of different measurements.
By using a Bayesian analysis, and by using a `non-informative' prior
for the `Hyper-Parameters', we find that for $M$ data sets  one 
should minimise 
\begin{equation}
-2 \; \ln P(\bfw| {\rm data} ) = 
\sum_{j=1}^{M} N_{j} \ln (\chi_{j}^{2}),   
\label{sum_hp}
\end{equation}  
where $N_j$ is the number of measurements in data set $j=1, ..., M$.
It is as easy to calculate this statistic as the standard $\chi^2$.
The corresponding HPs $\alpha_{{\rm eff},j} = N_j/\chi^2_j$ 
provide useful diagnostics on the reliability of different data sets. 
We emphasize that a low HP assigned to an experiment does not necessarily
mean that the experiment is `bad', but rather it calls attention to look 
for systematic effects or better modelleing. 

We have applied the HP analysis to a set of various CMB measurements
and estimated the Hubble constant $H_{0}$ 
(for a fixed flat CDM $\Omega_{\rm m} = 1 - \lambda = 0.3$ model).
While the standard $\chi^2$ approach 
gives a wide range for $H_0$, the Hyper-Parameter analysis
suggests two distinct values of $H_0$, $\sim 50$ and $\sim 70$
km/sec/Mpc. 
It remains to be understood why the ensemble of CMB experiments tends to give
two competing values (quite common in the history of the
Hubble constant !).
It would be most interesting to see how these values 
change when combined with other cosmological probes
(and their corresponding HPs).

   In estimating  $H_0$
    we have assumed that the other cosmological parameters 
    (like $\Omega_{\rm{m}}$ and
    $\Omega_\Lambda$) were known and, consequently, such an estimate is
    conditional on the assumed values of these parameters. This is not 
    strictly necessary, actually, since we may obtain a more general
    estimation of $H_0$ by marginalising over $\Omega_{\rm{m}}$, 
$\lambda$, 
    and the other cosmological parameters, in a way similar to what we have
    done with the HPs. 
One may also generalise the above method for more specific applications.
Two aspects which can be modified according to specific  problems 
are the priors $P(\alpha_j)$ and the probability functions
$P(D_j|\bfw)$. We shall discuss elsewhere these extensions in application 
to various cosmological probes.





\bigskip
\bigskip

{\bf Acknowledgments:} 
We thank G. Rocha for her contribution to the data analysis, 
and A. Dekel and I. Zehavi  
for helpful discussions.
SLB acknowledges the receipt of a PPARC studentship. 
OL thanks the IAG (S\~ao Paulo) for  hospitality.
LSJ thanks FAPESP, CNPq and
PRONEX/FINEP for the support to his work.


\end{document}